\documentclass[twocolumn,showpacs,preprintnumbers,amsmath,amssymb]{revtex4}

\usepackage{graphicx}
\usepackage{dcolumn} 
\usepackage{bm}      

\usepackage{textcomp}

\begin{document}

\title{Impact of elasticity on the piezoresponse of adjacent
ferroelectric domains investigated by scanning force microscopy}

\author{Tobias Jungk}
\email{jungk@uni-bonn.de}
\author{\'{A}kos Hoffmann}
\author{Elisabeth Soergel}

\affiliation{Institute of Physics, University of Bonn,
Wegelerstra\ss e 8, 53115 Bonn, Germany\\}

\date{\today}

\begin{abstract}
As a consequence of elasticity, mechanical deformations of crystals
occur on a length scale comparable to their thickness. This is
exemplified by applying a homogeneous electric field to a
multi-domain ferroelectric crystal: as one domain is expanding the
adjacent ones are contracting, leading to clamping at the domain
boundaries. The piezomechanically driven surface corrugation of
micron-sized domain patterns in thick crystals using large-area top
electrodes is thus drastically suppressed, barely accessible by
means of piezoresponse force microscopy.
\end{abstract}

\maketitle

\section{Introduction}

Ferroelectric domain patterns are intensively investigated due to
their increasing practical importance, e.g.\ for frequency
conversion~\cite{Fej92}, electrically controlled optical
elements~\cite{Eas01}, photonic crystals~\cite{Bro00} or
high-density data storage~\cite{Cho05}. For their characterization a
visualization technique with high lateral resolution is required.
Among a wealth of techniques~\cite{Soe05} piezoresponse force
microscopy (PFM) has become a standard tool for visualizing
micron-sized domain structures. For PFM a scanning force microscope
is operated in contact mode with an alternating voltage applied to
the tip. In ferroelectric samples this voltage causes thickness
changes via the converse piezoelectric effect~\cite{New} and
therefore vibrations of the surface which lead to oscillations of
the cantilever that can be read out with a lock-in
amplifier~\cite{Jun06,Alexe}. Quantitative analysis of PFM images is
complicated by the strongly inhomogeneous electric field generated
by the tip~\cite{Jun07a}. A possibility to overcome this difficulty
consists in the application of large-area electrodes on the sample
surfaces, thereby generating a plane plate capacitor-like electric
field configuration inside the sample.

PFM with large-area top electrodes has been realized on piezoceramic
thin films~\cite{Auc97,Gru98,Gru03}. The lateral resolution was not
observed to be affected by this electrode configuration. This is
plausible as mechanical coupling between adjacent grains in ceramics
can be assumed to be weak. Therefore they can deform independently
according to their crystallographic orientation and thus their
piezoelectric tensor elements. Indeed, there is no report so far on
PFM measurements of differently orientated domains within a single
grain covered by a metal layer. This, however, corresponds to the
situation of a metallized multi-domain single crystal.

\begin{figure}[ttt]
\includegraphics{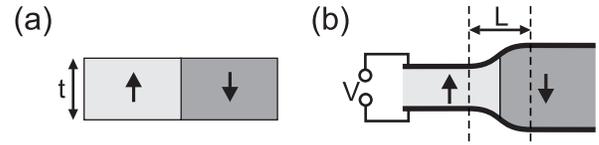}
\caption{\label{fig:Jungk1}
{(a) Bi-domain crystal of thickness $t$ and its deformation (b)
taking place on the length scale $L$ when applying a homogeneous
electric field to it. Note that the width of the domain wall itself
is not affected by the deformation of the crystal.}}
\end{figure}

Unfortunately, there is no analytical solution to the problem of
crystal deformation at a domain boundary in a homogeneous electric
field taking clamping into account. However, it can be estimated
from elasticity theory~\cite{Tim} and from finite element
calculations~\cite{Jac04} that clamping affects the surface
distortion on a length scale on the same order of magnitude as the
thickness of the crystal. This is exemplified in
Fig.~\ref{fig:Jungk1} where the deformation of a bi-domain crystal
of thickness $t$ in a homogeneous electric field is shown
schematically.

In this contribution we investigate the impact of elasticity on the
surface corrugation of multi domain samples. Due to their wide
applicability we used lithium niobate ($\rm LiNbO_3$) crystals
exhibiting 180$^{\circ}$ domains only. The width of the domain walls
is expected to be a few unit cells~\cite{Mey02}; from
high-resolution electron microscopy images they are known to be
narrower than 3\,nm \cite{Foe99}. When imaging domain walls with
PFM, however, they appear to be wider due to the limited resolution
of PFM, typically in the order of 50\,nm~\cite{Jun07b}.

\section{Experimental Procedure}

For the investigations we used a commercial scanning force
microscope (SMENA, NT-MDT), modified to allow the application of
voltages to the tip ($\rm 10\,V_{pp}$, $\sim 30\,$kHz). All PFM
images were recorded from the $X$-output of a dual-phase lock-in
amplifier (SRS 830), thus being unaffected by the background
inherent to PFM measurements~\cite{Jun07c}. The experiments were
carried out with $z$-cut $\rm LiNbO_3$ crystals (thicknesses:
500\,\textmu m and 80\,\textmu m). The crystals either had one
domain boundary or they were periodically poled (PPLN). We partly
metallized the samples with gold (Au) or copper (Cu) layers of 30 --
50\,nm thickness; one sample was covered with a Cu-layer of 6\,nm
thickness only. Structured metal layers were fabricated using
electron microscopy grids as masks for the evaporation. All samples
were grounded at the backside by a homogenous metal electrode and
mounted on a piezomechanically driven translation stage with a
travel range of 300\,\textmu m.

We used cantilevers from MikroMasch, some of them conductively
coated with Ti-Pt. To determine the quality of the electrical
connection between tip and metal layer, we applied an alternating
voltage to the tip and recorded the voltage at the layer via lock-in
amplification while scanning the sample. The obtained images show
the connectivity between tip and metal layer. For the non-coated,
highly n-doped silicon tips no electrical connection was found (most
probably because of a naturally growing oxide layer of some nm
thickness~\cite{Sze}). However, also for Ti-Pt coated tips,
electrical contact for both the Au- and the Cu-layers was not
reliable. This may be caused by the extremely small contact area.
Hence, if the application of voltages to the metal layer was
desired, we directly connected the voltage source with an extra
wire.

\section{Experimental Results}
\subsection{Clamping at a single domain wall}

\begin{figure}[ttt]
\includegraphics{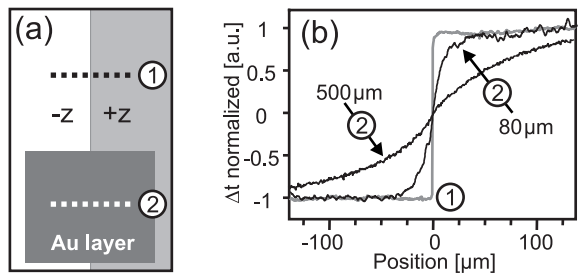}
\caption{\label{fig:Jungk2}
{(a) Diagram of the $\rm LiNbO_3$ samples used to determine the
surface distortion across the domain boundary with and without an
additional Au-layer. (b) Line scans recorded at the positions
indicated in (a) perpendicular to the domain boundary for two
samples of 500\,\textmu m and 80\,\textmu m thickness. At the
non-covered area the line scans $\textcircled 1$ coincide for both
crystals. $\Delta t$: normalized surface deformation of the
sample.}}
\end{figure}

To demonstrate the effect of clamping between adjacent 180$^\circ$
domains, we partly metallized two bi-domain crystals of thicknesses
500\,\textmu m and 80\,\textmu m with a 50\,nm Au-layer on an area
of several mm$^2$ (Fig.~\ref{fig:Jungk2}(a)). Then we measured the
length of the deformation of the surface across the domain boundary
with $\textcircled 2$ and without $\textcircled 1$ the Au-layer. In
Fig.~\ref{fig:Jungk2}(b) line scans of 260\,\textmu m length at the
positions indicated in (a) are shown. For better comparability the
line scans are normalized to the same maximum crystal deformation.
As the full deformation for the 500\,nm thick sample was not
achieved within the scan range, we measured the crystal response
some mm away from the domain boundary and fitted the line scan with
a modified hyperbolic tangent to estimate the width of the surface
deformation~\cite{Jun07b}. Independent of the thickness of the
sample, the surface distortion for both crystals using standard PFM
(with the tip acting as electrode) show step-like profiles
$\textcircled 1$. On the metallized region, however, the deformation
of the crystal reaches its maximum at a distance of $\sim
300$\,\textmu m away from the domain boundary for the 500\,\textmu m
thick crystal and at $\sim 35$\,\textmu m for the 80\,\textmu m
thick crystal, respectively. The surface deformation across the
domain boundary is thus of the same order of magnitude as the
thicknesses of the crystals. This is what can be expected from
elasticity theory due to clamping between adjacent
domains~\cite{Tim}.

\subsection{Impact of clamping on PFM imaging}

\begin{figure}[ttt]
\includegraphics{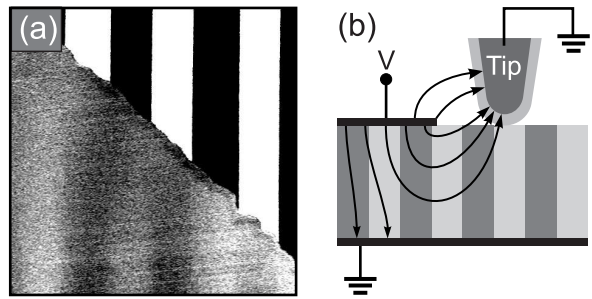}
\caption{\label{fig:Jungk3}
{(a) PFM image of PPLN partially covered with a conducting (30\,nm
thick) Cu-layer. The alternating voltage was applied to the
Cu-layer, the tip and the back-electrode being grounded (b). The
contrast in (a) is enhanced to reveal the surface corrugation under
the electrode of $\sim 2$\,pm amplitude. The image size is $25\times
25$\,\textmu m$^2$.}}
\end{figure}

To investigate the impact of clamping on the surface corrugation of
multi-domain crystals, we partly metallized a 500\,\textmu m thick
PPLN crystal (period $\Lambda=8$\,\textmu m) with a 30\,nm thick
Cu-layer. Figure~\ref{fig:Jungk3} shows a recorded PFM image with
the corresponding sample configuration and electrical connection
scheme. To avoid a short circuit between the grounded tip and the
Cu-layer, we used a non-coated silicon tip, exhibiting a few
nm-thick insulating oxide layer. The PFM measurements
(Fig.~\ref{fig:Jungk3}(a)) revealed a piezomechanically driven
surface deformation of 140\,pm in the non-metallized part of the
sample. However, underneath the Cu-layer the surface corrugation was
measured to be~$< 2$\,pm.
The latter result can be compared to the slope of the line scan for
the 500\,\textmu m thick crystal in Fig.~\ref{fig:Jungk2}(b) which
was found to be $\sim 0.4$\,pm/\textmu m at the domain boundary. The
distance between neighboring domain boundaries in this PPLN crystal
is 4\,\textmu m. Therefore a surface corrugation of $\sim 1.6$\,pm
is expected which matches fairly well the measured data. This result
is consistent with the assumption that the surface corrugation is
reduced by a factor of $\Lambda /2\,t$ for periodically clamped
samples, with $t$ being the thickness of the sample~\cite{Ste98}.
The shape of the surface grating in the homogeneous electric field
was found to be sinusoidal.

Furthermore, it is striking that the contrast in the two parts of
the image is inverted. This becomes evident, however, when looking
at the direction of the electric field inside the crystal
(Fig.~\ref{fig:Jungk3}(b)). Analyzing carefully the image, it can
also be observed that near the edge of the Cu-layer the contrast is
reduced even more. This can be explained as follows: when the tip is
at the edge of the layer, the crystal is not only clamped because of
the PPLN structure underneath but in addition the non-covered part
of the crystal remains undeformed because of the absence of the
electrical field.

\subsection{Optical diffraction experiments}

\begin{figure}[ttt]
\includegraphics{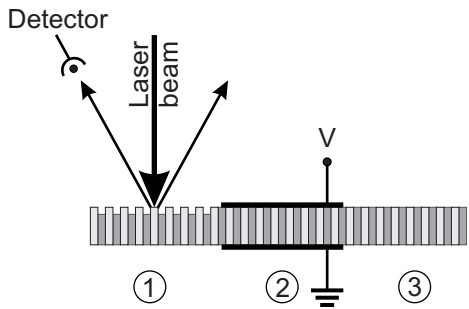}
\caption{\label{fig:Jungk4}
{Schematics of the PPLN sample used for the optical diffraction
experiments. In section $\textcircled 1$ a distinct surface relief
grating was obtained by etching with hydruofloric acid, in the
central section $\textcircled 2$ Au-electrodes allow the application
of up to 1\,kV to the sample, and section $\textcircled 3$ remained
unchanged to perform the experiments with liquid electrodes.}}
\end{figure}

To sustain these results and to exclude possible artifacts in the
PFM measurements, we performed optical diffraction
experiments~\cite{Ste98,Cap03} with the surface-relief grating.
Therefore an identical PPLN sample ($\Lambda~=~8$\,\textmu m,
thickness 500\,\textmu m) was prepared as follows
(Fig.~\ref{fig:Jungk4}): one part of the sample was etched with
hydrofluoric acid to reveal the domain structure as a distinct
surface-relief grating, one part of the sample was covered with a
50\,nm thick Au-electrodes and the rest of the sample remained
unchanged.

In a first step the exact read-out angle for the diffraction was
determined with the etched part of the sample. Then the sample was
moved with a translation stage in order to irradiate the part
covered with the Au-electrodes. Although applying 1\,kV to them, we
could not observe any diffracted laser beam. We repeated the
experiment with the sample mounted in a special holder with liquid
electrodes~\cite{Mue03}, applying 10\,kV to it. In this setup the
unmodified part of the sample was investigated, thereby excluding
any mechanical restriction of the Au-layer. However, also in this
case no diffraction from the piezomechanically induced
surface-relief grating was observed. According to the piezoelectric
coefficient $d_{33}=8.1$\,pm/V~\cite{Jaz02} without considering
clamping, a surface corrugation of $\sim 80$\,nm should emerge when
applying 10\,kV to the crystal, easily detectable by optical means.
We attribute the missing diffraction the extremely smoothed surface
relief grating caused by clamping. This is in agreement with the
results from our PFM measurements, performed with a PPLN crystal
with same period $\Lambda$: Clamping reduces the piezomechanically
driven surface corrugation by a factor of $\sim 70$. Therefore,
although applying 10\,kV to the crystal, the expected surface
modulation is only $\sim 1$\,nm. The detection of such small surface
corrugations, however, requires a more sophisticated
setup~\cite{Ste98} and is beyond the scope of our experimental
setup.

\subsection{Influence of metallic islands on PFM imaging}

\begin{figure}[ttt]
\includegraphics{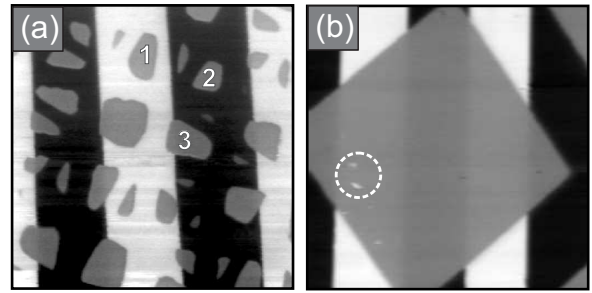}
\caption{\label{fig:Jungk5}
{Effect of isolated metal layers on PFM imaging. (a)~PPLN crystal
covered with Cu-islands (40\,nm thickness). The size of the islands
varies from 2\,\textmu m$^2$ to 70\,\textmu m$^2$. Some of them are
located on a positive (1) or on a negative domain face only (2) and
some lie across a domain boundary (3). The image size is $50\times
50$\,\textmu m$^2$. (b)~PFM image of a PPLN crystal partly covered
with a 6\,nm thick Cu-layer. Inside the circle the Cu-layer has
defects. The image size is $60\times 60$\,\textmu m$^2$, the full
$z$-range of the images is 140\,pm.}}
\end{figure}

Finally, we want to underline the importance of a reliable
electrical connection to the metal layer. Thus, we performed a
series of experiments avoiding any connection deliberately. In this
situation the metallization acts like a shield and therefore no
electric field penetrates into the crystal. Consequently no PFM
signal at all can be detected under the electrode. This is shown in
Fig.~\ref{fig:Jungk5}(a) where 40\,nm thick \textmu m-sized
Cu-islands have been deposited on top of a PPLN crystal. Some of the
islands are placed at one domain face only (1 and 2), whereas others
lie across a domain boundary (3). Independent on their position
relative to domain boundaries no contrast can be detected under the
metallic islands. The missing contrast has nothing to do with a size
effect of the electrode and therefore a changed field distribution
inside the crystal~\cite{Jun07a}. Note that the PFM signal from the
islands is the exact median of the two PFM signals from the $+z$ and
the $-z$ domain faces. This is consistent with the system-inherent
background described previously~\cite{Jun06}.

To give further evidence that the missing contrast at the Cu-islands
in Fig.~\ref{fig:Jungk5}(a) is due to a shielding mechanism, we
evaporated a very thin Cu-layer (6\,nm) on top of a PPLN crystal.
Such thin metal layers are known to be not fully
conducting~\cite{Jac90}. Thus, the shielding by the metal layer is
not perfect any more, that is why we observed a faint contrast of
the domain pattern underneath the Cu-layer
(Fig.~\ref{fig:Jungk5}(b)). In comparison to the gratings generated
in a homogeneous electric field this grating shows a pronounced
step-like profile and its amplitude is roughly 7\% of the amplitude
measured at the non-coated area. Note that at some positions, e.g.\
the ones indicated with the circle in Fig.~\ref{fig:Jungk5}(b) the
Cu-layer is damaged and therefore the full PFM signal can be
observed.

\section{Conclusions}

We have investigated the consequences of elasticity on the clamping
between adjacent ferroelectric domains. Thereby we could confirm
theoretical predictions on the length scale clamping affects the
surface deformation and its dependence on the thickness of the
crystal. The effects being too small to be detected by optical
means, we used piezoresponse force microscopy (PFM) capable of
measuring deformations of only a few picometers. The obtained
results have also a direct impact for the technique of PFM itself,
quantifying the drawback for the resolution using a top metal layer
when investigating multi-domain crystals.

\section*{Acknowledgments}


We thank Michael K\"{o}sters for the optical measurements and the
poling of the bi-domain samples. Many thanks also to Boris Sturman
for helpful discussions. Financial support of the DFG research unit
557 and of the Deutsche Telekom AG is gratefully acknowledged.

\end{document}